\documentclass[onecolumn,
               superscriptaddress,
               floatfix,
               longbibliography, 
               showkeys,apl]{revtex4-2}
\usepackage[utf8]{inputenc}
\usepackage{graphicx} 
\usepackage{bm}	
\usepackage{color}                     
\usepackage{xcolor}
\usepackage{epsfig}
\usepackage{amsmath} 
\usepackage{amssymb} 
\usepackage{bbold}
\usepackage[T1]{fontenc}

\usepackage{float}
\usepackage{mathtools}
\usepackage{xparse}
\usepackage{hyperref}
\usepackage{chemformula}
\usepackage{physics}
\usepackage{subfiles}
\usepackage{placeins}

\usepackage{tcolorbox}
\tcbuselibrary{breakable,xparse,skins}
\definecolor{bg}{gray}{0.95}

\usepackage[margin=1.25in]{geometry}

\usepackage[caption=false]{subfig}

\begin{document}

\title{Connecting the Hamiltonian structure to the QAOA energy and Fourier landscape structure}


\author{Michał Stęchły} \email{michal.stechly@gmail.com}
\affiliation{Zapata Computing Canada Inc., 25 Adelaide St E, Toronto, ON, M5C3A1}

\author{Lanruo Gao}
\affiliation{Zapata Computing Canada Inc., 25 Adelaide St E, Toronto, ON, M5C3A1}

\author{Boniface Yogendran}
\affiliation{Zapata Computing Canada Inc., 25 Adelaide St E, Toronto, ON, M5C3A1}

\author{Enrico Fontana}
\affiliation{University of Strathclyde, 16 Richmond St, Glasgow G1 1XQ, UK}
\affiliation{National Physical Laboratory,  Hampton Rd, Teddington TW11 0LW, UK}

\author{Manuel S. Rudolph}
\affiliation{Zapata Computing Canada Inc., 25 Adelaide St E, Toronto, ON, M5C3A1}
\affiliation{Institute of Physics, Ecole Polytechnique Fédérale de Lausanne (EPFL), CH-1015 Lausanne, Switzerland}


\date{\today}



\begin{abstract}

In this paper, we aim to expand the understanding of the relationship between the composition of the Hamiltonian in the Quantum Approximate Optimization Algorithm (QAOA) and the corresponding cost landscape characteristics. QAOA is a prominent example of a Variational Quantum Algorithm (VQA), which is most commonly used for combinatorial optimization. The success of QAOA heavily relies on parameter optimization, which is a great challenge, especially on scarce noisy quantum hardware. Thus understanding the cost function landscape can aid in designing better optimization heuristics and therefore potentially provide eventual value. We consider the case of 1-layer QAOA for Hamiltonians with up to 5-local terms and up to 20 qubits. In addition to visualizing the cost landscapes, we calculate their Fourier transform to study the relationship with the structure of the Hamiltonians from a complementary perspective. Furthermore, we introduce metrics to quantify the roughness of the landscape, which provide valuable insights into the nature of high-dimensional parametrized landscapes. While these techniques allow us to elucidate the role of Hamiltonian structure, order of the terms and their coefficients on the roughness of the optimization landscape, we also find that predicting the intricate landscapes of VQAs from first principles is very challenging and unlikely to be feasible in general.

\end{abstract}



\maketitle

\section{Introduction}\label{Section:introduction}
\textit{Variational Quantum Algorithms} (VQAs) are a family of hybrid quantum-classical algorithms designed for \textit{near-term intermediate scale quantum} (NISQ) \cite{Preskill2018quantumcomputingin} devices. The goal of a VQA is to prepare a quantum state that optimizes a given cost function (commonly also called \textit{objective function} or \textit{loss function}).
A cost function can be, for example, the expectation value of a hermitian operator such as an Hamiltonian~\cite{peruzzo2014variational}, or the distance between generated and target distributions of measurement outcomes~\cite{benedetti2019generative}. Unfortunately, it has become evident that the highly non-convex cost landscapes are hard to reliably navigate-- especially with noise incurred from finite measurement statistics. The particular shape of each cost landscape and its interaction with the chosen optimization algorithm are key factors that determine the reliability and efficiency of a VQA~\cite{cerezo2020variational}. 
Numerous studies (such as Refs.~\cite{PECT,Metalearning, Kubler2020adaptiveoptimizer}) that go beyond random parameter initialization and optimization naive gradient descent were published in the past years to cleverly explore the parametrized cost landscapes, but a universal solution has, perhaps unsurprisingly, not yet been found. We argue that it is essential to understand and appreciate each particular family of cost landscapes defined by problem instances of interest in order to select an informed parameter initialization strategy and an appropriate optimization algorithm.

The \textit{Quantum Approximate Optimization Algorithm} (QAOA)~\cite{Farhi_QAOA} is a type of VQA, which is typically used to solve combinatorial optimization problems. This is done by approximately finding the ground state of a diagonal Ising Hamiltonian which encodes the problem ~\cite{Lucas2014}. In QAOA, the cost function is usually the expectation value of the problem Hamiltonian on a parametrized state, i.e., its energy. The circuit ansatz for the quantum state is highly structured and problem dependent-- a property that QAOA does not share with many types of VQAs. As such, there are significantly fewer design choices that go into utilizing QAOA compared to more generic VQA implementations. One of the remaining being the choice of the optimizer for the particular problem instance.

\begin{figure}
    \centering
    \includegraphics[width=0.9\linewidth]{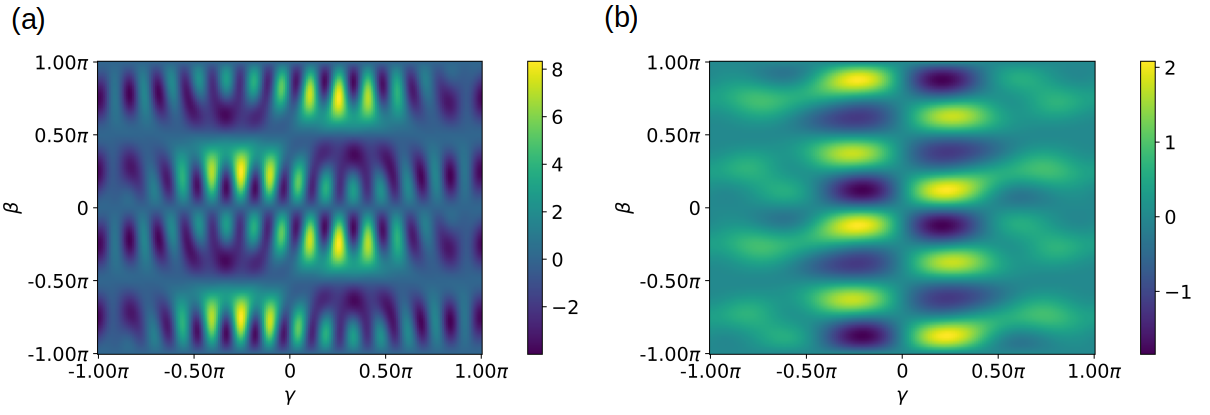}
    \caption{Cost landscape scans of the Hamiltonians $H_1$ (a) and $H_2$ (b) with $p=1$ layer QAOA. The scans visually demonstrate that even though $H_1$ has an apparently simpler structure than $H_2$, its cost landscapes exhibit many more local minima which are likely to lead to difficult optimization.  $H_1$ and $H_2$ are defined in Eqs.~\eqref{eq:H1} and~\eqref{eq:H2}, respectively.}
    \label{fig:H1_H2}
\end{figure}
\FloatBarrier

While working on analyzing the \textit{Variational Quantum Factoring} algorithm (VQF)~\cite{VQF}, we became aware of the following pair of Ising Hamiltonians, which challenged our understanding of QAOA landscapes:

\begin{equation}\label{eq:H1}
    H_1 = -2.75 Z_0 - 3.25 Z_1 +3.75 Z_0 Z_1 
\end{equation}

\begin{equation}\label{eq:H2}
\begin{aligned}
    H_2 = &\frac14 \sum_{i=0}^5 Z_i + \frac34 \sum_{i = \{0,2,4 \}} Z_i Z_{i+1} + \frac18\sum_{i= \{ 0,2\}} Z_iZ_{i+2}Z_4Z_5  \\ 
    + &\frac18\sum_{i=2}^5 (-1)^{i+1} Z_0Z_i + \frac18\sum_{i=2}^5 (-1)^i Z_1Z_i + \frac18\sum_{i=2}^3 (-1) ^i Z_iZ_4 + \frac18\sum_{i=2}^3 (-1) ^{i+1} Z_iZ_5 \\
    - &\frac18\sum_{i=2}^5 Z_0Z_1Z_i - \frac18\sum_{i=0}^3 Z_iZ_4Z_5 - \frac18\sum_{i = \{0,1,4, 5 \}} Z_2Z_3Z_i,
\end{aligned}
\end{equation}
where $Z_{i}$ denote Pauli $Z$ operators acting on qubit $i$. These Hamiltonians were the initial motivation for this work. 
One may naively assume that the Hamiltonian with three 1- and 2-body terms has a significantly simpler QAOA cost landscape than one with 36 up to 4-body terms. However, after visually analyzing the cost landscapes of these Hamiltonians, we discovered that this assumption is in fact incorrect (see Fig.~\ref{fig:H1_H2}). This led to the desire to pin down and understand which properties of the Hamiltonian influence the shape of its cost landscape and to what extent.


This work aims to study how the structure of a Hamiltonian influences the resulting QAOA cost landscape, and by this contributes to our general understanding.
Our approach is based on visualization of the cost landscape, but additionally, utilization of their Fourier transform, i.e., the Fourier landscape, as a key tool.
In order to more quantitatively evaluate the measured landscapes, we introduce roughness metrics on the cost and Fourier landscapes which can be correlated with how ``simple'' or ``complicated'' both landscapes appear. We demonstrate the predictability of certain features of the landscapes by systematically manipulating the QAOA Hamiltonians.
While it is unlikely that these or similar roughness metrics can be used in general to predict which optimizer will be most successful in minimizing the cost, they may be valuable to study a particular family of QAOA instances of interest.

Section~\ref{Section:Background} recapitulates the QAOA method, how it can be studied from the perspective of the Fourier spectrum, and introduces the roughness metrics used throughout this work. In Sec.~\ref{Section:Results} we present our findings on how the particular form of a Hamiltonian affects the QAOA cost and Fourier landscapes. Finally, we conclude in Sec.~\ref{Section:Conclusion} and point out possible future research directions.



\section{Background}\label{Section:Background}

\subsection{QAOA and its cost function}\label{background:QAOA_and_cost_function}

QAOA is a type of VQA that aims to find the ground state of an Ising Hamiltonian. The QAOA ansatz, i.e., the architecture of the circuit, is defined by the digitized time evolution under the problem Hamiltonian (cost Hamiltonian) as well as a second mixer Hamiltonian. It consists of a total of $p$ cost Hamiltonian layers (parametrized by parameters $\gamma \in \mathbb{R}^p$) and mixer layers (parametrized by parameters $\beta \in \mathbb{R}^p$), respectively, where $p$ is a user-controlled hyperparameter which controls the expressive power of the circuit. As such, the circuit ansatz is highly problem-specific and only sparsely parametrized by $2p$ parameters.

In this work, we restrict ourselves to studying QAOA with $p=1$, as well as the mixer Hamiltonian $\sum_{i=0}^{n} X_i$, which follows the commonly-used original QAOA ansatz~\cite{Farhi_QAOA}. Here $n$ is the number of qubits and $X_i$ are Pauli X operators.  In this form, and with a 2-body problem Hamiltonian, the cost function value $\mathcal{C}(\beta, \gamma)$, i.e., the energy, has a closed-form expression which was defined in Ref.~\cite{Ozaeta_2022}. Such a Hamiltonian can be interpreted as an undirected graph with its vertices $ V = {1, ..., n}$ and edges $E \subset V \times V$, where each vertex has an associated bias $h_i$ and each edge has coupling strength $J_{ij}$. The Hamiltonian consequently consists of the sum of 1- and 2-body terms ($C_i$ and $C_{ij}$, respectively)
\begin{equation}
    H = \sum_{i\in V} C_i + \sum_{(i,j) \in E} C_{ij},\;\; C_i := h_i Z_i,\; C_{ij} := J_{ij} Z_i Z_j \; ,
\end{equation}
where the notation $(i, j) \in E$ denotes neighboring vertices on the graph that are connected by an edge.

The cost function can be written as
\begin{equation}\label{eq:QAOA_cost_function}
    \mathcal{C}(\beta, \gamma):=\sum_{i \in V}\left\langle C_i\right\rangle+\sum_{(i, j) \in E}\left\langle C_{i j}\right\rangle.
\end{equation}
The closed-form formulas for the individual cost terms have been derived in \cite{Ozaeta_2022} and are then given by:
\begin{align}
    \left\langle C_i\right\rangle 
        &= h_i \sin (2 \beta) \sin \left(2 \gamma h_i\right) \prod_{(i, k) \in E} \cos \left(2 \gamma J_{i k}\right) \label{eq:Ci}\\
    \left\langle C_{i j}\right\rangle
        &= \frac{J_{i j} \sin (4 \beta)}{2} \sin \left(2 \gamma J_{i j}\right) \biggl[\cos \left(2 \gamma h_i\right) \prod_{\substack{(i, k) \in E \\ k \neq j}} \cos \left(2 \gamma J_{i k}\right) + \cos \left(2 \gamma h_j\right) \prod_{\substack{(j, k) \in E \\ k \neq i}} \cos \left(2 \gamma J_{j k}\right)\biggr]  \label{eq:Cij}  \nonumber \\
        &-\frac{J_{i j}}{2}(\sin (2 \beta))^2 \prod_{\substack{(i, k) \in E \\
        (j, k) \notin E}} \cos \left(2 \gamma J_{i k}\right) \prod_{\substack{(j, k) \in E \\
        (i, k) \notin E}} \cos \left(2 \gamma J_{j k}\right) \nonumber \\
        & \times \biggl[\cos \left(2 \gamma\left(h_i+h_j\right)\right) \prod_{\substack{(i, k) \in E \\
        (j, k) \in E}} \cos \left(2 \gamma\left(J_{i k}+J_{j k}\right)\right) \nonumber \\
        & -\cos \left(2 \gamma\left(h_i-h_j\right)\right) \prod_{\substack{(i, k) \in E  \\
        (j, k) \in E}} \cos \left(2 \gamma\left(J_{i k}-J_{j k}\right)\right) \biggr] 
\end{align}

By inspecting Eqs.~\eqref{eq:Ci} and ~\eqref{eq:Cij}, one can notice that all terms are trigonometric functions whose frequencies depend on the parameters $h_i$ and $J_{ij}$ of the problem Hamiltonian. Thus one can immediately derive properties of the landscape such as its maximum frequency and parameter symmetries like parity. Unfortunately, it applies only to Hamiltonians with 1- and 2-body terms, as closed-form formulas for higher-order terms are not known. Furthermore, as we elaborate in the Sec.~\ref{sec:fourier}, the trigonometric form of QAOA cost functions justifies the use of Fourier analysis to characterize their landscape.

\subsection{Fourier analysis of the landscape}
\label{sec:fourier}
The Fourier transform is the quintessential classical signal processing tool. It can be used to isolate the sinusoidal components of any given function. In the particular case of VQAs, the cost function is evaluated on a \textit{parameterized quantum circuit} (PQC): $\mathcal{C}(\theta) = \langle 0 | U^{\dag} (\theta) HU(\theta)|0\rangle$, where $H$ is the cost Hamiltonian, and $\theta = (\theta_1, ..., \theta_N)$ the parameters that control the variable gates in the quantum circuit. Typically, these gates are generated by Hermitian operators.
It has been demonstrated multiple times across the literature that the cost functions of quantum algorithms can be described as Fourier series (see e.g., Refs.~\cite{schuld2021effect, vidal2018calculus, gil2020input, nakanishi2020sequential, parrish2019jacobi, ostaszewski2021structure, wierichs2022general, Enrico_1, Enrico_2}).
In order words, we can write
\begin{equation}
    \mathcal{C}(\theta) = \sum_{\omega \in \Omega} c_\omega e^{i\omega\cdot\theta}
\end{equation}
where the coefficients $c_\omega$ above can be recovered by performing a $N$-dimensional Fourier transform
\begin{equation}\label{eq:main-ft}
	c_{\omega} = \dfrac{1}{(\sqrt{2\pi})^N}\underset{[0,2\pi]^N}{\int} e^{-i \omega\cdot\theta} \mathcal{C}(\theta)  \,d^N\theta\,,
\end{equation}
where $\omega \cdot \theta$ is an inner product.
The frequencies $\omega\in\Omega\subset\mathbb{R}^N$  depend only on the Hamiltonians that generate the parameterized unitary operations. Suppose we have a single gate $e^{-i\theta M}$ generated by a Hermitian operator $M$, $N=1$. Then the corresponding frequencies of the cost function obey
\begin{equation}
    \omega \in \{\lambda_i(M) - \lambda_j(M)\}_{i,j}\, ,
\end{equation}
meaning all possible differences between pairs of eigenvalues $\lambda_i(M)$ of $M$~\cite{schuld2021effect}.
Conversely, the values of the coefficients $c_{\omega}\in\mathbb{C}$ depend on the circuit, the input state and the cost Hamiltonian $H$~\cite{vidal2018calculus, schuld2021effect,Enrico_1}, and in general are hard to estimate classically.
In the case of $p=1$ QAOA, $N=2$, the parameters are $\theta = (\beta, \gamma) $ and $\omega = (f_{\beta}, f_{\gamma})$ are the corresponding frequencies. Since the unitary parameterized by $\gamma$ is generated by the problem Hamiltonian $H$, the frequencies $f_\gamma$ will depend on the differences between the eigenvalues of $H$.

As outlined in Refs.~\cite{Enrico_1,Enrico_2}, the Fourier coefficients contain information about general features of the cost landscape. Most relevant for our discussion, in Ref.~\cite{Enrico_2} it was shown that the number of nonzero Fourier coefficients, also known as the Fourier sparsity, is an important metric for optimization. For instance, it was shown that narrow gorges (an obstacle to optimization) imply many nonzero coefficients. Roughly speaking, this occurs because a narrow gorge is a concentration of the cost function in parameter space, and by the uncertainty principle of Fourier analysis this implies a frequency spectrum that is spread out.
Conversely, the same paper showed that a very sparse spectrum means that the cost function be efficiently reconstructed classically via compressed sensing, thus making its optimization much simpler.

In summary, Fourier analysis provides us with a powerful and suitable tool for characterizing the cost landscape of QAOA instances.




\subsection{Roughness metrics} \label{sec:roughness-metrics}

Humans have an intuitive understanding of what is meant by a ``rough'' cost function landscape. Upon comparing the two cost landscapes in Fig.~\ref{fig:H1_H2}, one may conclude that the landscape in panel (a) is ``rougher'' than the one in panel (b). This conclusion, however, is subjective and not quantifiable. Furthermore, simple inspection becomes problematic for cost landscapes in higher dimensions (i.e., with more parameters), which are significantly harder to inspect and prone to the existence of artifacts due to the choice of visualization technique.

One potential precise definition of roughness could be the density of local optima (minima or maxima depending on the algorithm) and saddle points in a given landscape. If the landscape presents numerous suboptimal local optima, not only will it appear rough to a human observer, but optimization algorithms will likely struggle to find good solutions. 
Both gradient-based and gradient-free optimizers have the tendency to remain trapped in local optima instead of reaching global optima, but the exact behavior depends on the particular optimization algorithm of choice. Therefore, we seek objective metrics that are able to quantify roughness, which can be fed back into the decision which optimizer to use. To this end, various metrics were explored. Here we present two selected roughness metrics that we have found to be useful, which correlate well and yet are complimentary in certain aspects. The advantages and disadvantages of the other metrics we considered are presented in Appendix~\ref{secA2}.

\subsubsection{Metric 1: Total variation}\label{sec:background_total_variance}

Total variation (TV) has previously been used to estimate roughness in neural network loss landscapes~\cite{Wu2021UnderstandingLL}. Its name refers to the fact that it captures how much a function varies across a finite domain. It can be calculated by integrating the absolute value of the gradient: 
\begin{equation}\label{eq:tv}
	TV(f)= \int_{a}^{b} \lvert f'(x)\rvert \,dx\
\end{equation}
If a function oscillates more over a fixed domain, then it will have a steeper average slope. TV thus represents a good measure for the oscillatory behavior of a function, which makes it a good candidate for a roughness index. 
In our implementation we normalize the expression in Eq.~\eqref{eq:tv} maximum range of function values
\begin{equation}\label{eq:tv-f-def}
[f] = \max\limits_{a\leq x\leq b}f(x)-\min\limits_{a\leq x\leq b}f(x)\, .
\end{equation}
This gives a normalized TV
\begin{equation}\label{eq:tv-normalized}
TV_{\text{norm}}(f)= \frac{1}{[f]}\int_{a}^{b} \lvert f'(x)\rvert \,dx \, ,
\end{equation}
which is invariant to scaling of the cost function ($f \rightarrow \alpha f$), $\alpha\in\mathbb{R}$.
Because one does in practice not have access to a closed-form expression for the function gradient, we use a first-order numerical integration approximation of the TV which utilizes a first-order numerical gradient approximation on a grid of parameters. The TV then takes the simple form
\begin{equation} \label{eq:tv-discrete}
TV_{\text{discrete}}(f)=\frac{1}{[f]}\sum_{i=1}^{m}\left| f\left( a+(i+1)r\right)-f\left( a+ir\right) \right| \, ,
\end{equation}
where $r=\frac{b-a}{m}$ and $m$ is the number of steps in the discrete integral.

To generalize to multivariate functions, we take 1D slices of the landscape by picking random vectors originating from a common point and sample the landscape along those directions. The normalized TV is then calculated for each section with the mean value representing the final roughness index.
We measure along 200 random directions with $m=200$ samples along each of these directions. This value provided a good compromise between computation time and the reliability of the results (for more information, see Appendix~\ref{secA2}).
For each section, the size of the sampling domain $[a, b]$ is set such that it covers exactly one period of the cost function.

\subsubsection{Metric 2: Fourier density}
\label{sec:background_fourier_density}

As discussed in Sec. \ref{sec:fourier}, the frequency modes revealed by the Fourier transform offers a natural perspective to study quantum landscapes. 
In particular, in Ref.~\cite{Enrico_2} the Fourier sparsity defined as the number of nonzero Fourier coefficients was shown to be an important metric for the purpose of optimization. Mathematically, if $\textbf{c}_\omega$ is the vector of Fourier coefficients, the Fourier sparsity is defined as its 0-norm $\|\textbf{c}_\omega\|_0$.

In this work, we adapt Fourier sparsity to become a metric of landscape roughness. 
We will refer to the metric by the name \textit{Fourier density} (FD) to avoid the potential confusion in interpreting phrases such as ``large Fourier sparsity'' (which should describe that Fourier coefficients are not sparse).
However, this metric as defined above through the 0-norm poses a challenge for numerical implementations such as ours, as the finite precision of floating point calculations may lead to an artificially large value. One potential solution is to set a threshold, below which a coefficient is considered to be zero and is not counted. This, however, introduces an arbitrary free parameter.


In Ref.~\cite{unknown_sparsity}, the numerical sparsity 
\begin{equation}\label{eq:fourier_density}
    s(\textbf{c}_\omega) := \|\textbf{c}_\omega\|_1^2 / \|\textbf{c}_\omega\|_2^2
\end{equation}
was proposed as a solution to the numerical instability of the 0-norm, where $\|\textbf{c}_\omega\|_1^2$ and $\|\textbf{c}_\omega\|_2^2$ are the 1- and 2-norm, respectively. $s(\textbf{c}_\omega)$ is in fact a sharp lower bound on $\|\textbf{c}_\omega\|_0$ with equality if and only if all non-zero coefficients are of equal magnitude. Because of its lack of arbitrary threshold and its numerical stability, we utilize the expression in Eq.~\eqref{eq:fourier_density} as the second roughness metric and we refer to it as \textit{Fourier Density}. Other variants of this metric that we considered are discussed in Appendix ~\ref{secA2}.

\section{Results}\label{Section:Results}
For visualization purposes, we use the \textit{orqviz} package~\cite{orqviz}, which is a Python library for visualizing the cost function landscapes in parametrized algorithms. For details on the visualization techniques see Appendix~\ref{sec:viz-methods}.

All the code needed to reproduce the results of this paper can be found at \url{https://github.com/Boniface316/qaoa_landscape}.

\subsection{Toy problem}\label{subsec_results:toy}

We begin with the study of a family of 2-qubit Hamiltonians, \begin{equation}\label{eq:toy_hamiltonian}
    H(a,b,c) = aZ_0 + bZ_1 + cZ_0Z_1 \, .
\end{equation} 
The limited number of degrees of freedom allows for a closed-form analytical expression of the cost function, which in turn can be compared to the empirically measured cost and Fourier landscapes. 
The cost function for this Hamiltonian can be derived from Eqs.~\eqref{eq:Ci} and~\eqref{eq:Cij} and reads
\begin{equation} \label{eq:toy-problem}
\begin{aligned}
\mathcal{C}(\beta, \gamma) &= \langle C_0 \rangle + \langle C_1 \rangle + \langle C_{01} \rangle \\
& = a \sin(2 \beta) \sin(2 a \gamma) \cos(2 c \gamma) + b \sin(2 \beta) \sin(2 b \gamma) \cos(2 c \gamma)  \\ 
& \hspace{0.5cm}  + \frac{c}{2} \left[ \sin(4 \beta) \sin(2 c \gamma) (\cos(2 a \gamma) + \cos(2 b \gamma)) \right. \\
\ &  \left. \hspace{0.5cm} - \sin^2(2\beta) ( \cos(2 (a + b) \gamma) - \cos(2  (a-b)\gamma))  \right]
.
\end{aligned}
\end{equation}
Upon first examination, we see that each coefficient $a$, $b$ and $c$ affects both the amplitudes and the frequencies of all partaking terms. The existence of the two-qubit term in the Hamiltonian gives rise to cosine terms with frequencies that are proportional to either the sum or the difference of the coefficients $a$ and $b$. This already hints at the slew of frequencies that contribute to larger problems with more coefficients. 
This form of the cost function can be directly linked to the Fourier spectrum of $\mathcal{C}$, as $\beta$ and $\gamma$ are present only as arguments of the trigonometric functions. As described in Sec.~\ref{sec:fourier}, the Fourier transform allows us to define $f_{\beta}$ and $f_{\gamma}$ for every cost term. The formulation in Eq.~\eqref{eq:toy-problem}, however, allows us to observe that all the $\sin$ terms with $\beta$ in the argument have frequencies that do not depend on the values of the coefficients. This implies that $f_{\beta}$ will be independent of the particular choice of Hamiltonian coefficients.

\begin{figure}[!htb]
    \centering
    \includegraphics[width = 0.99\linewidth]{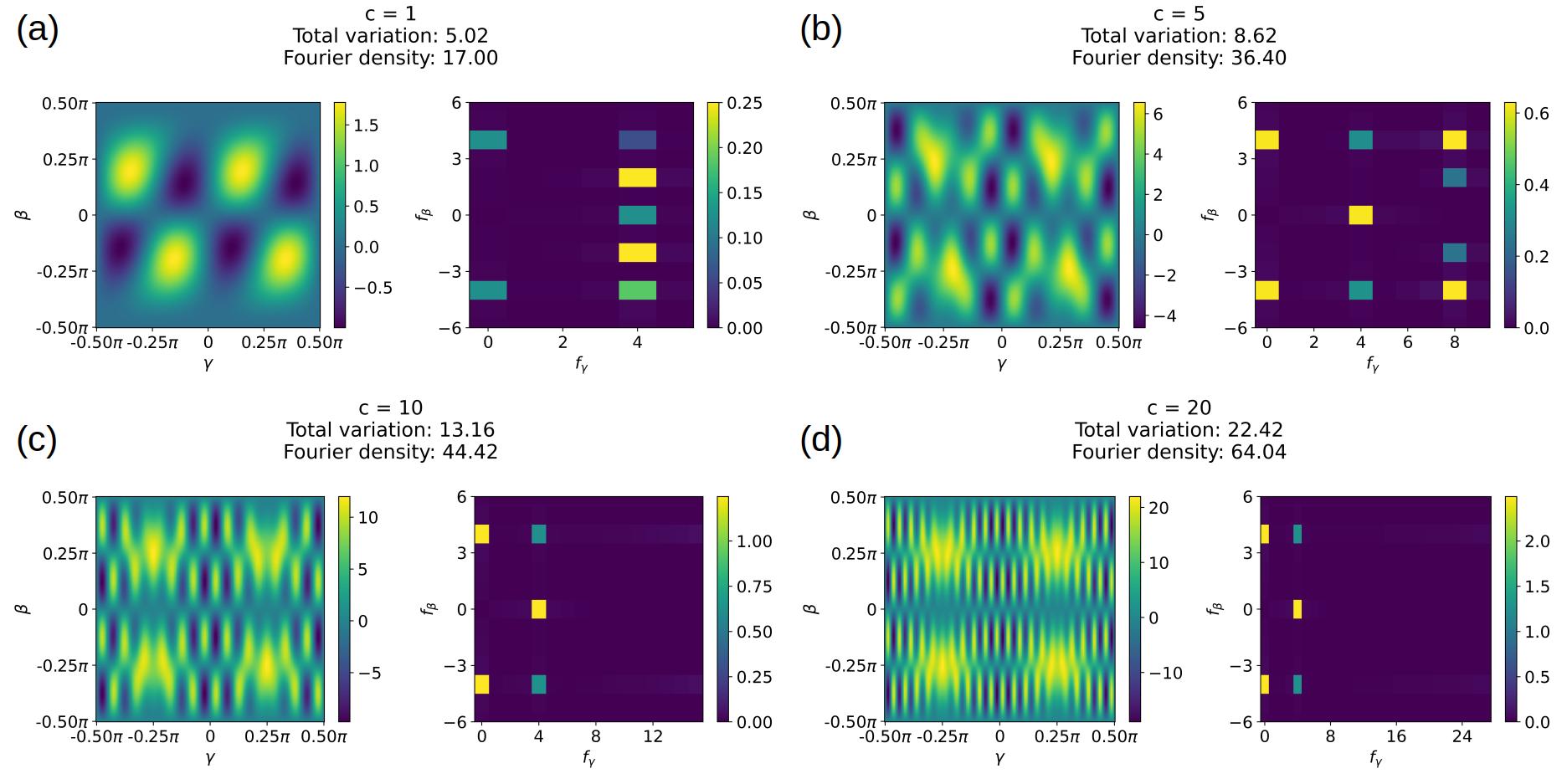}
\caption{Cost and Fourier landscapes scans for the Hamiltonians ${H(c) = Z_0 + Z_1 + c Z_0Z_1}$, with $c = \{1, 5, 10, 20\}$. The analytically calculated frequencies are shown in \mbox{Table}~\ref{table:toy-problem-frequencies} and match the frequencies observed in the numerical experiments up to finite-size artifacts.
}
\label{fig:phase_1_varying_c}
\end{figure}

To further simplify the problem, we first assign $a=b=1$ and calculate expected frequencies for various values of $c$. Detailed calculations can be found in Appendix~\ref{secA1}, the expected outcomes are presented in Table~\ref{table:toy-problem-frequencies}, and the empirically observed cost and Fourier landscapes are depicted in Fig.~\ref{fig:phase_1_varying_c}. Because there are terms that depend on the value of $c$ and others that do not, as $c$ is increased some frequencies shift up, whereas others do not change. This can also be observed in the cost landscape scans, where some underlying features of the landscape stay consistent, while higher frequency contributions are superimposed and differ in each case. Visually, one would say that the cost landscapes with higher-frequency contributions are rougher-- a trend that is confirmed by the roughness metrics.

Note that in Table~\ref{table:toy-problem-frequencies} we actually list the absolute value of the frequencies. In reality, the Fourier transform, as defined in Eq.~\ref{eq:main-ft}, will output both positive and negative-valued frequencies. 
This is because each trigonometric term gives rise to an equal number of positive and negative frequency terms with the same absolute frequency values, e.g. $\cos(4\beta) = \frac{1}{2}(e^{i4\beta} + e^{-i4\beta})$. Note that, while the \textit{frequencies} $\omega$ are the same in absolute value, the \textit{coefficients} $c_\omega$ at negative frequencies may have a completely different value from those at positive ones. As a simple calculation shows, the cost function being real implies that $c_{-\omega} = c^*_{\omega}$, but for the case of more than one parameter (where $\omega$ is a vector), this still leaves some freedom in the negative frequency coefficients. Indeed, take our case with 2 parameters $\theta = (\beta, \gamma)$ and frequencies $\omega = (f_\beta, f_\gamma)$. The fact that the cost function is real-valued implies that $c_{(f_\beta, f_\gamma)} = c^*_{(-f_\beta, -f_\gamma)}$. In other words, the Fourier landscape is point-symmetric around the origin. 
Hence we depict the negative frequencies in the Fourier spectrum for the $f_\beta$ direction, only cutting the spectrum in half once to account for the symmetry $c_{-\omega} = c^*_{\omega}$.

\begin{center}
\begin{table}[H]
\centering
\begin{tabular}{|l|l|l|}
\hline
c  &  Frequencies from linear terms &  Frequencies from quadratic term                              \\ \hline
   & (2c$\pm$2, 2)                 & (2c$\pm$2, 4) (4, 4) (4, 0) (0, 4)                    \\ \hline
1  & (4, 2)                        & (4, 4) (4, 0) (0, 4)                    \\ \hline
5  & (8, 2) (12, 2)                & (8, 4) (12, 4) (4, 4) (4, 0) (0, 4) \\ \hline
10 & (18, 2) (22, 2)               & (18, 4) (22, 4) (4, 4) (4, 0) (0, 4)                  \\ \hline
20 & (38, 2) (42, 2)                & (38, 4) (42, 4) (4, 4) (4, 0) (0, 4)                  \\ \hline
\end{tabular}
\caption{Frequencies for Hamiltonian of form $H(c) = Z_0 + Z_1 + c Z_0 Z_1$ (see Eq.~\eqref{eq:toy_hamiltonian}), written as $(|f_{\gamma}|, |f_{\beta}|)$. When multiple terms with the same frequencies would be present in a particular row, their amplitudes will be added together. This summation results in a reduced number of points in the Fourier plot. The constant term $(0, 0)$ is omitted.
Appendix \ref{secA1} contains detailed steps on how these values have been obtained and why the term $(0, 2)$ for $c=1$ is missing.
}
\label{table:toy-problem-frequencies}
\end{table}
\end{center}
\FloatBarrier

\FloatBarrier

We used the Hamiltonian in Eq.~\ref{eq:toy_hamiltonian} also to study the effect of non-integer coefficients. As explained in Appendix~\ref{sec:vis_fourier}, if the \textit{greatest common divisor} (GCD) of the coefficients is small, the period of the cost function can become very large.
For instance, if $a=0.171$, $b=0.340$, $c=0$, the GCD is 0.001 and the period of the resulting cost function would be $\frac{\pi}{0.001} = 1000 \pi$. 
If we round the coefficients to $a=0.17$, $b=0.34$ and $c=0$ the period would be equal to $\frac{\pi}{0.17}$. However, we can also note that this Hamiltonian is equivalent to one with $a=1$, $b=2$, and $c=0$, up to a constant re-scaling factor in both frequency and amplitude, which would further simplify the analysis. Since the GCD has been drastically increased from 0.001 to 0.17, and consequently the period is reduced by orders of magnitude, one might expect that the cost landscape is significantly different. 
Interestingly, we show that this is not the case.

\medskip

\begin{figure}[!htb]
    \centering
\includegraphics[width = 0.99\linewidth]{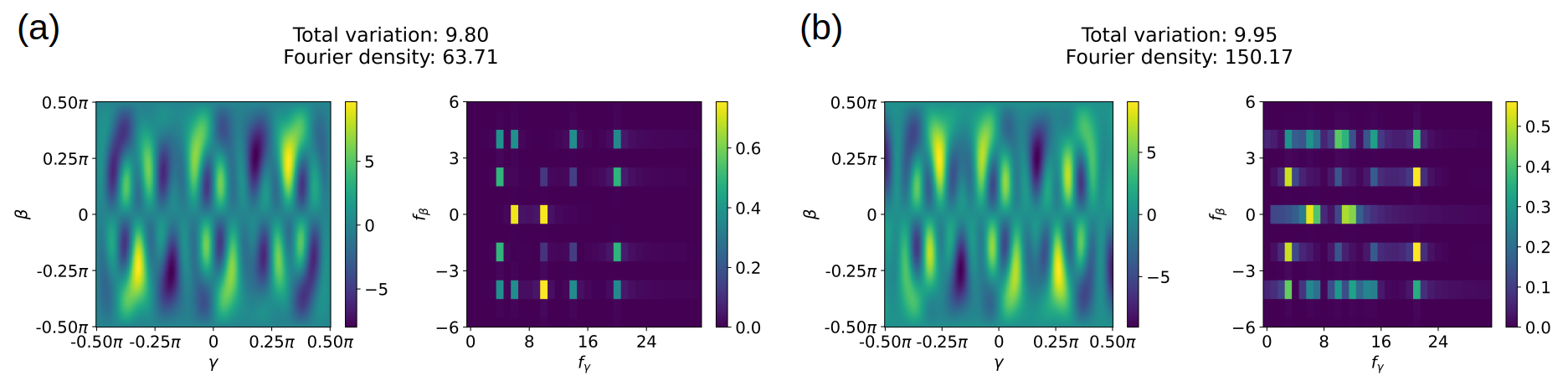}
\caption{Cost and Fourier landscapes for two similar Hamiltonians with integer coefficients (a) and non-integer coefficients (b). The Hamiltonians are ${H_a = 1 Z_1 + 4 Z_2 + 6 Z_1 Z_2}$ and ${H_b = 1.23 Z_1 + 4.47 Z_2 + 6.05 Z_1 Z_2}$, respectively. While the cost landscapes appear to be very similar, the Fourier spectra differ more strongly. This is a result of the fact that in discrete Fourier transformation, non-integer frequencies do not result in well-localized peaks, but rather in distributions of frequencies.}
\label{fig:phase_2_non_integer}
\end{figure}

We generated multiple Hamiltonians with random coefficients, where the coefficients were drawn from a uniform distribution ranging from -10 to 10. For each of these Hamiltonians, a corresponding Hamiltonian was constructed with coefficients rounded to the nearest integers. The cost landscapes of these Hamiltonians were analyzed visually and through the use of roughness metrics.
We observe a high degree of similarity among the majority of the cost landscapes generated between the non-integer and their corresponding rounded integer-coefficient Hamiltonian. Naturally, the differences in the cost landscapes are most noticeable when the rounding errors are high. A representative example can be seen in Fig.~\ref{fig:phase_2_non_integer}.
On the flip side, the Fourier spectrum is significantly different between the pairs of Hamiltonian.
By nature of the discrete Fourier transform of the cost landscape, the frequencies ($f_\beta, f_\gamma$) are integer-valued. Therefore, if the Hamiltonian eigenvalues are integer-valued, also their sum and difference result in integers. This results in well-localized and distinct frequencies in the Fourier landscape. In contrast, the original un-rounded Hamiltonian with non-integer coefficients exhibits non-integer frequencies which need to be distributed over adjacent integer frequencies. 
This is a phenomenon which is captured well by the Fourier density metric, as it is highly sensitive to the presence of additional frequencies in the spectrum. However, one can argue that these are simply artifacts of the discrete Fourier transform. It follows that the Fourier density should not be used to compare the roughness between landscapes where one is integer-valued and the other is non-integer-valued. Consequently, we constrain our study to Hamiltonians that do not fall into this case.


\FloatBarrier

\subsection{Higher k-body terms}\label{ssec:higher-k}

Here we analyze the effect of the presence of $k$-body terms in the Hamiltonian on the cost and Fourier landscapes in QAOA. Hamiltonians with $k \geq 3$ body terms are rarely studied throughout the literature. One apparent explanation is that such Hamiltonians do not natively map to problems represented as graphs, which are typical problem instances for the QAOA algorithm. However, we believe it is vital to understand the effect that such terms have on the optimization landscape, as they are present in certain non-graph-related problem instances (e.g., the VQF algorithm~\cite{VQF}).

\begin{figure}[!htb]
\centering
\includegraphics[width = 0.99\linewidth]{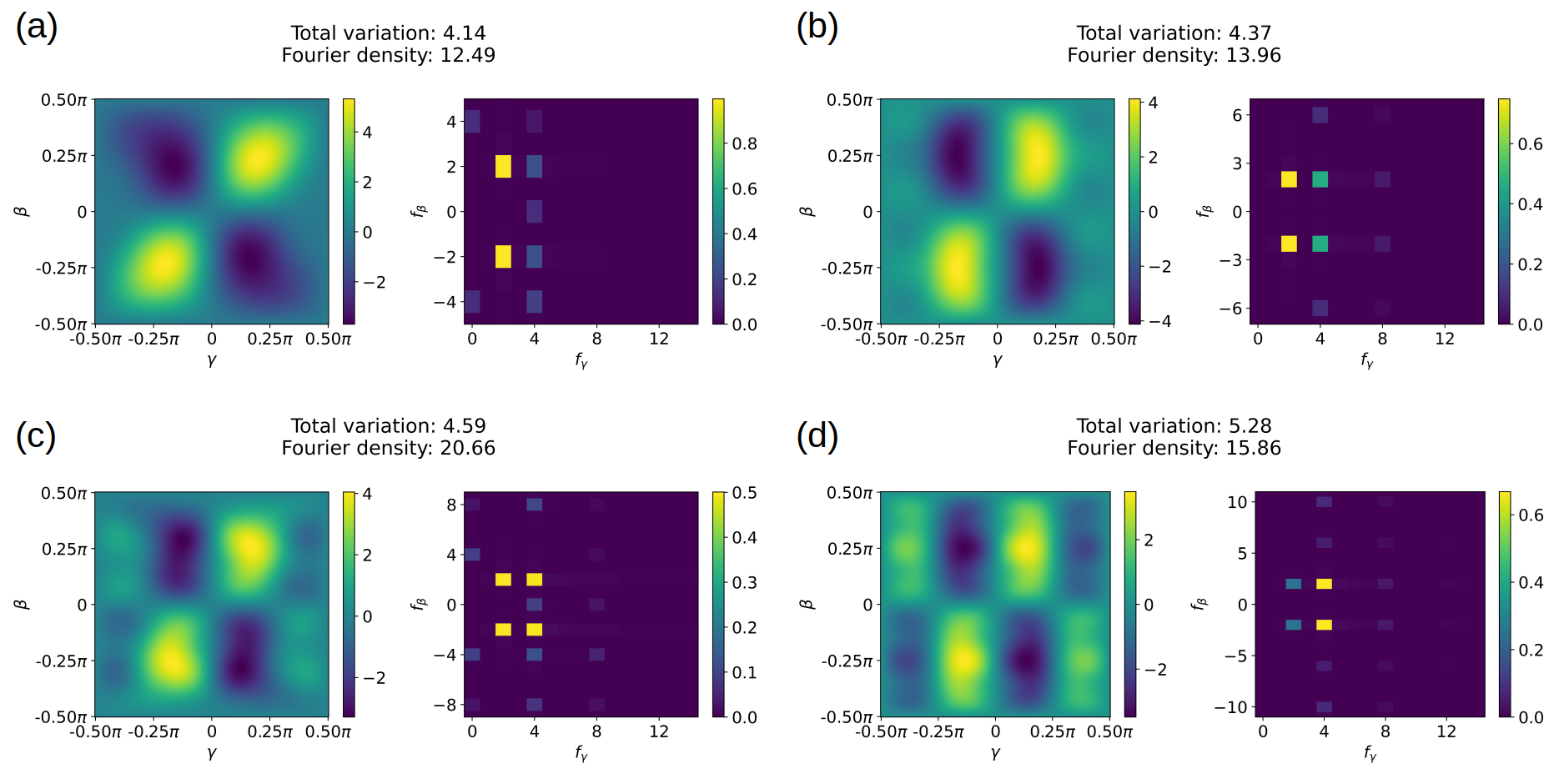}
\caption{Cost and Fourier landscapes for the Hamiltonians $\mathcal{H}_2$ (a), $\mathcal{H}_3$ (b), $\mathcal{H}_4$ (c) and $\mathcal{H}_5$ (d) in Eq.~\eqref{eq:k-local-Hs}. We notice that as higher-order terms are introduced, Total Variation tends to increase. The same clear trend is not reflected by the Fourier Density. We can also see the highest possible frequency $f_{\beta}$ increase as $\pm 2\times k^*$.
}
\label{fig:fig_4_A}
\end{figure}

To construct Hamiltonians for our experiments, we used as a base a 6-qubit Hamiltonian consisting of all $1$-body Pauli $Z$ terms:
\begin{equation}
    \mathcal{H}_1 = \sum_{i=1}^6 Z_i\ .
\end{equation}
To control for the number of terms, we add one $k^*=2,3,4,5$ body term to the Hamiltonian,
\begin{equation}\label{eq:k-local-Hs}
\begin{aligned}
    &\mathcal{H}_2 = \mathcal{H}_1 + Z_1 Z_2,  \quad\quad\quad\; \mathcal{H}_3 = \mathcal{H}_1 +  Z_1 Z_2 Z_3\\
    &\mathcal{H}_4 = \mathcal{H}_1 + Z_1 Z_2 Z_3 Z_4 ,  \quad \mathcal{H}_5 = \mathcal{H}_1 + Z_1 Z_2 Z_3 Z_4 Z_5 \ ,
\end{aligned}
\end{equation}
and study the resulting landscapes.
The results can be viewed in Fig.~\ref{fig:fig_4_A}. The cost landscapes clearly exhibit contributions of higher frequencies that are generated by the higher-order terms, and which lead to the emergence of additional local minima. The TV metric mirrors the observed trend and increases steadily with $k^*$, while the Fourier Density decreases in the last case with $k^*=5$.
This highlights that the latter metric does not necessarily correlate with the density of local minima in the cost landscape.

As it can be read off in the Fourier landscapes, higher order terms increase the number of terms in the Fourier spectrum, as well as their possible $f_{\beta}$ frequencies. The highest possible frequency in this dimension is equal to twice the order of the highest term, i.e. $\pm 2 \times k^*$. This outcome is consistent with our expectations based on the analysis of equations Eqs.~\eqref{eq:Ci} and~\eqref{eq:Cij}. However, as we can see in the Fourier landscapes, replacing the particular term also affects the relative weights of other Fourier coefficients.


\subsection{One large coefficient}
\label{ssec:one_large}
\begin{figure}[!htb]
        \centering
        \includegraphics[width = 0.99\linewidth]{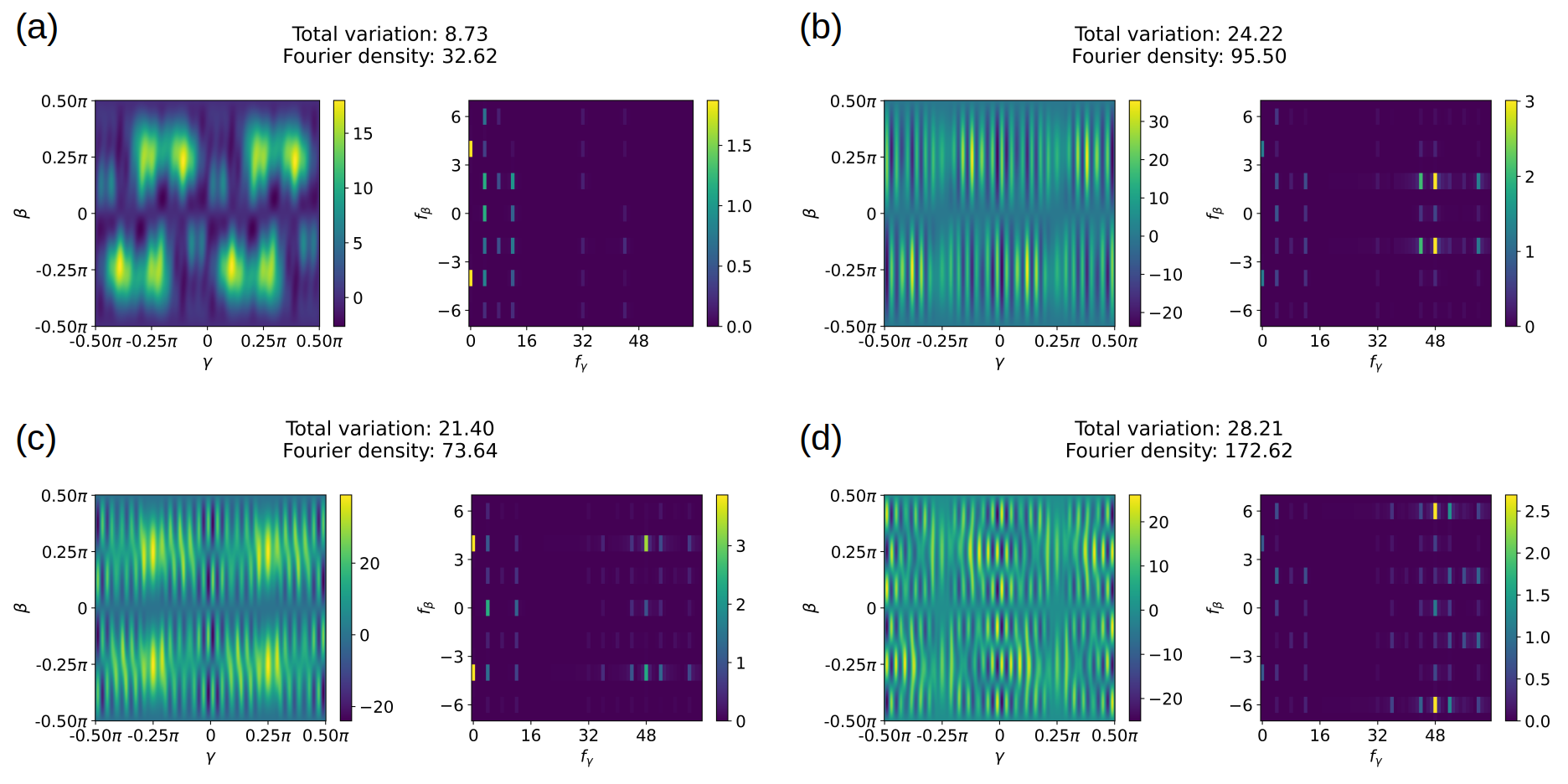}
\caption{Cost and Fourier landscapes for the Hamiltonians $\mathcal{H}_6$ (a), $\mathcal{H}_7$ (b), $\mathcal{H}_8$ (c) and $\mathcal{H}_9$ (d) which are a variation of the Hamiltonian in Eq.~\ref{eq:H123}. For the latter three Hamiltonians, one $k$-body term, i.e., $k^*=1, 2, 3 \, ,$ respectively, had its coefficient increased to $25$. 
}
\label{fig:single_big}
\end{figure}

In Sec.~\ref{subsec_results:toy}, we show that the magnitude of Hamiltonian coefficients can drastically impact the roughness of the cost landscape-- an effect that we have observed repeatedly throughout our studies.  On the other hand, we show in Sec.~\ref{ssec:higher-k} that introducing a single high-order term also tends to increase roughness. A natural question to ask is which of the two factors has the stronger impact? To study this, we use the Hamiltonian 
\begin{equation}\label{eq:H123}
    \mathcal{H}_{6} = \sum_{i=1}^6 Z_i + \sum_{i < j}^6 Z_iZ_j + \sum_{i < j < k}^6 Z_iZ_jZ_k
\end{equation}
as the base and create three new Hamiltonians from it. For each we pick the first term of the order of $k^*=1$ (the $Z_1$ term), $k^*=2$ (the $Z_1 Z_2$ term) or $k^*=3$ (the $Z_1 Z_2 Z_3$ term), and increase its coefficient to 25. We label them as $\mathcal{H}_7$, $\mathcal{H}_8$ and $\mathcal{H}_9$, respectively. Their cost and Fourier landscapes are shown in Fig.~\ref{fig:single_big}.

A first observation is that the cost landscapes in fact become significantly rougher when a single large coefficient is introduced. Furthermore, we observe the addition of new dominant frequency at $(f_\gamma, f_\beta) = (\pm 2\times 25, 2\times k^*)$. However, while TV and FD correlate well with each other, in this case we see that roughness metrics do not consistently increase with $k^*$. Overall this clearly suggests that the effect of a single (or a few) outlier coefficients can completely warp the cost landscape, but the interplay with the order of the term is unclear. The coefficient manipulates $f_\gamma$ frequencies and the order of the term influences $f_\beta$. Their interference is what finally appears to determine the exact shape of the cost and Fourier landscapes.


\FloatBarrier

\subsection{Structure of the Hamiltonian}
\label{ssec:ham_structure}

In order to better understand the influence of Hamiltonian structure on the cost landscape, we analyze cases where we incrementally change Hamiltonians by adding one term at a time. More specifically, we interpolate between two types of Hamiltonians and study where the landscapes change most in terms of roughness. \textit{A priori} it is not evident whether the largest changes would happen when the structure of the Hamiltonian is broken, i.e. when the first different terms are added, or during the middle part of the interpolation when there is arguably the least structure.

The Hamiltonians studied here form the sequence
\begin{equation}\label{eq:interpolation_Hs}
\begin{aligned}
    &\mathcal{H}_{a} = \sum_{i=1}^6 Z_i \\
   &\mathcal{H}_{a,1} = \mathcal{H}_a + Z_0Z_1Z_2 \\
   &\mathcal{H}_{a,2} = \mathcal{H}_a + Z_0Z_1Z_2 + Z_0Z_1Z_3 \\
   &\quad \quad \quad \vdots\\
   &\mathcal{H}_{a,20} = \mathcal{H}_a + Z_0Z_1Z_2 + Z_0Z_1Z_3 + \cdots + Z_3Z_4Z_5 = \mathcal{H}_{b} \ ,
\end{aligned}
\end{equation}
and thus represent the interpolation between $\mathcal{H}_{a}$, the Hamiltonian containing all $k=1$ terms, and $\mathcal{H}_{b}$, the Hamiltonian containing all $k=1$ and $k=3$ terms.

\begin{figure}
        \centering
        \includegraphics[width = 0.99\linewidth]{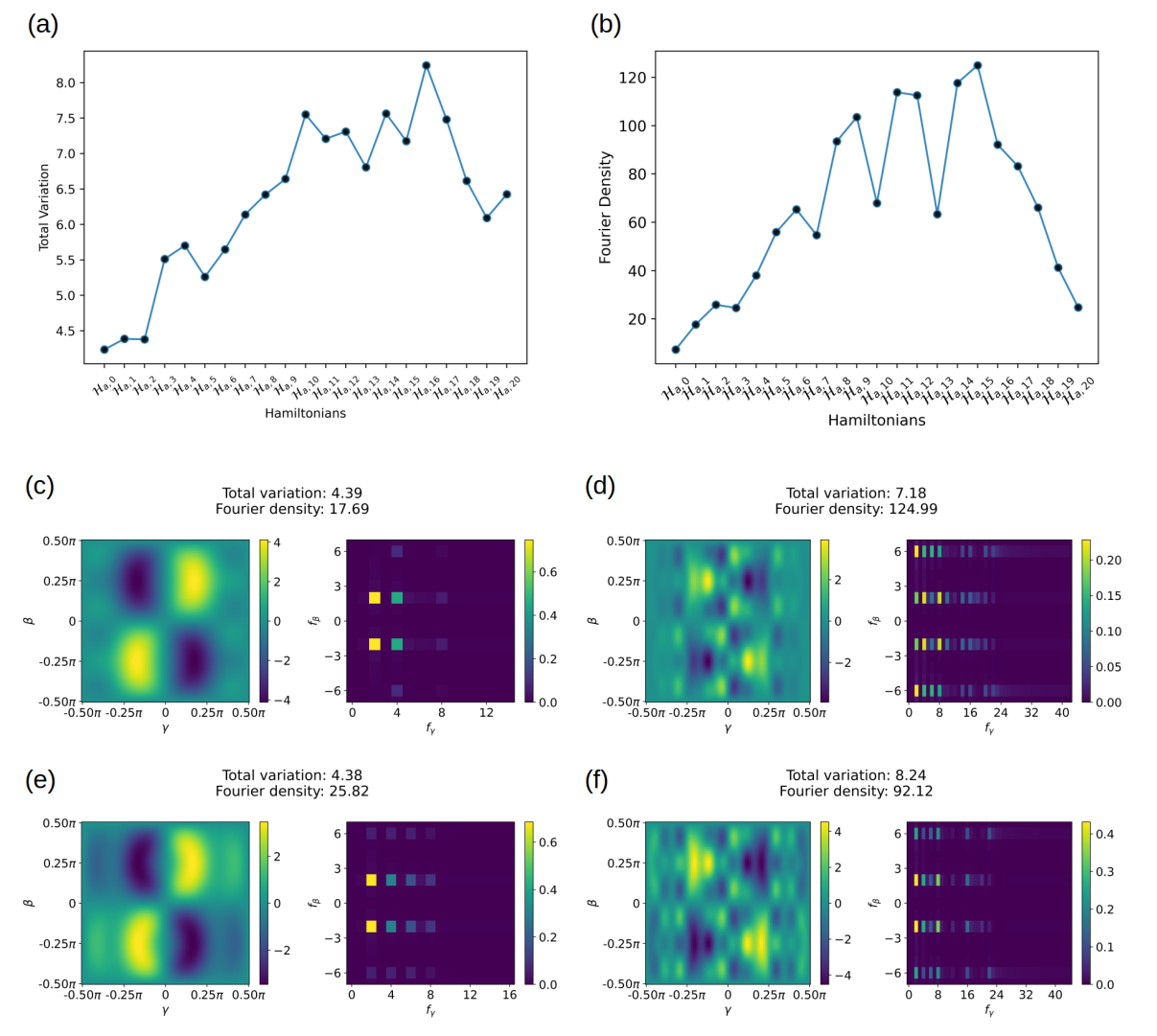}
\caption{Recorded roughness metrics (a, b) while interpolating between the Hamiltonians $\mathcal{H}_a$ and $\mathcal{H}_b$ by adding one term at a time (see Eq.~\eqref{eq:interpolation_Hs}). We additionally show two pairs of cost Hamiltonians with small and big differences in TV. Panels c) and e) show $\mathcal{H}_{a,1}$ and $\mathcal{H}_{a,2}$, whereas panels d) and f) show $\mathcal{H}_{a,15}$ and $\mathcal{H}_{a,16}$. The results demonstrate that adding or removing individual terms does not have a drastic influence on the landscapes when the dominant structure is mostly preserved. In contrast, large changes in the landscapes occur when there is not one particular dominant structure.}
\label{fig:fig_6}
\end{figure}

Panel a) and b) of Fig.~\ref{fig:fig_6} show the TV and FD values along the interpolation path where new terms are added one by one. These results imply that adding a single term can have both minor or significant effects on the landscapes depending on the current Hamiltonian. In particular, changes in roughness are largest and most erratic in the intermediate region between $\mathcal{H}_{a}$ and $\mathcal{H}_{b}$ where the landscapes also tend to be rougher overall.  
The fact that roughness is higher for the Hamiltonian $\mathcal{H}_b$, which contains all $k=3$ terms, is consistent with our observations in Sec.~\ref{ssec:higher-k} that Hamiltonians containing higher-order terms tend to exhibit rougher landscapes.
In panels c) - f) we present representative cost and Fourier landscape examples for the smallest and largest changes in roughness metrics.

Our findings suggest that predominant structures in the Hamiltonian (or when applicable, its graph representation) stabilize the cost landscape. In this regime, individual terms do not appear to contribute significantly, and it requires a significant shift in structure for the landscape to qualitatively change. We note that, as the interpolation can be understood in both directions, adding or removing individual terms has equivalent effects. 


\subsection{Parameter concentration and barren plateaus}\label{sec:concentration_and_bps}

The phenomenon in QAOA that is most known for applying Hamiltonians with particular structures is that of \textit{parameter concentration}.
It was shown in Ref.~\cite{farhi_param_concentration} that, for MaxCut problems on random 3-regular graphs of the same size and across graph sizes, local minima and maxima are located at similar parameter values. Among other works, this finding was underpinned in Ref.~\cite{orqviz}, where it was shown that in those cases the entire cost landscapes are very similar. It was also shown that the phenomenon is not just limited to the MaxCut problem or 3-regular graphs, and in fact persists for other types of problems and graphs, even when the coefficients are selected from a random distribution.

\medskip

In this work, we analyze the phenomenon of parameter concentration from the perspective of the cost and Fourier landscapes.
For this purpose, we study MaxCut problems on 3-regular graphs, which are graphs where each node is connected to exactly three other nodes.  
For a graph $G$ that is specified by the sets of $n$ nodes $V(G)$ and edges $E(G)$, we label a node as $v_i \in V(G)$, $i = 1,2,\dots n$ and the edge connecting nodes $v_i$ and $v_j$ is labeled as $(i,j)$. Each edge can have an individual weight $w_{ij}$ associated with it. The objective of the MaxCut problem is to find a partition of $V(G)$ into two sets $A$ and $B$, such that the sum of weights on the edges between $A$ and $B$ is maximized. 
We can encode whether a given node is in set $A$ or $B$ using a binary string $z\in\{-1, 1\}^{\otimes n}$, where $z_i=-1$ if node $i$ is in $A$ and $z_i=1$ if it is in $B$. The MaxCut cost function can then be written as the sum of weighted edges crossing $A$ and $B$, 

\begin{equation}
    \mathcal{C} = \sum_{(i,j) \in E(G)} \frac{1}{2} w_{ij} (1 - z_i z_j)\,.
\end{equation}
This cost function can naturally be expressed as the expectation value of a quantum state on the Hamiltonian
\begin{equation}\label{eq:maxcut_hamiltonian}
    \mathcal{H}_{\text{MaxCut}} = \sum_{(i,j) \in E(G)} \frac{1}{2} w_{ij}(\mathbb{1} - Z_i Z_j)\,,
\end{equation}
where $\mathbb{1}$ is the identity operator and $Z_iZ_j$ are Pauli $Z$ operators acting on qubits $i$ and $j$.
For our simulations, we removed the constant offset terms from the Hamiltonian that stem from the weighted identity operators.

\begin{figure}
    \centering
    \includegraphics[width=0.99\linewidth]{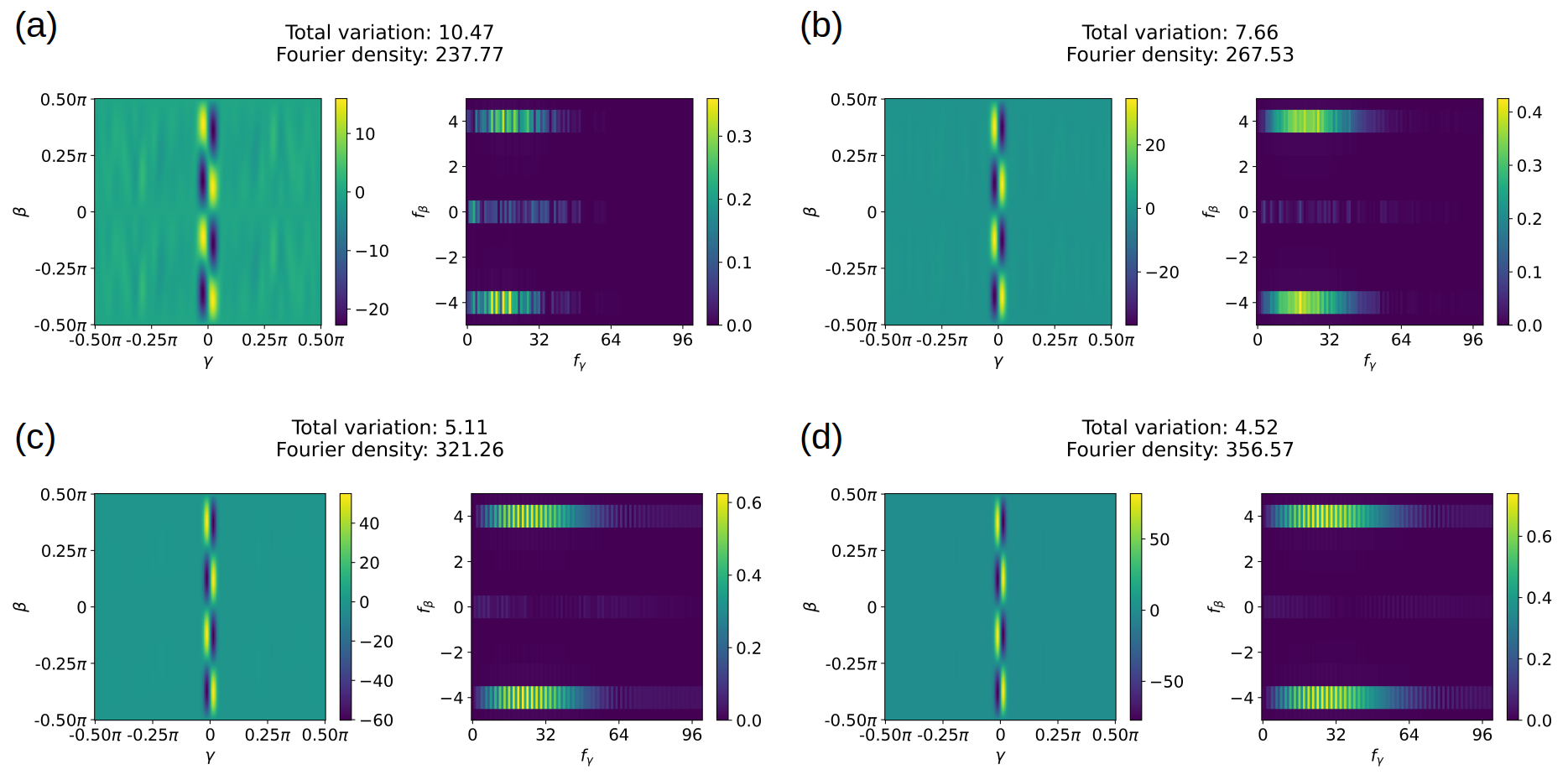}

\caption{Cost and Fourier landscapes of Hamiltonians for the MaxCut problem on random 3-regular graphs. The system sizes are $8$ (a), $12$ (b), $16$ (c), and $20$ (d) qubits. As the size of the problem increases, the cost and Fourier landscapes concentrate. We can see the onset of barren plateaus in the cost landscapes and the formation of an approximately bell-shaped distribution of the frequencies in the Fourier landscape. 
}

\label{fig:parameter_concentration}
\end{figure}

The features of the cost landscapes and their dependence on the values of the Hamiltonian coefficients are extensively discussed in Ref.~\cite{Ozaeta_2022}. In our study, we draw random 3-regular graphs of sizes $n=8, 12, 16$ and $20$, and with random uniform edge weights in the domain $w_{ij}\in[-10, 10]$. Representative results are shown in Fig.~\ref{fig:parameter_concentration}. 
In all cases, both the cost and Fourier landscape concentrate. It may be surprising that randomly connected graphs with a 3-regular structure and random weights concentrate in this manner, but our observations are robust across many repetitions, coefficient ranges, other types of random graphs (e.g., Erdős–Rényi graphs), and in agreement with the findings in Ref.~\cite{orqviz}.

In the Fourier picture, the coefficients with $f_{\beta}=0$ tend to get smaller as the size of the system increases, while the ones with $f_{\beta}=2$ form an approximately bell-shaped distribution with progressively widening spread. 
The cancellation of most of the frequencies also leads to a decrease in the value of the TV as the size of the problem increases, while values of Fourier Density steadily increase.

One interpretation of the cost landscape scans and the decreasing TV is the presence of barren plateaus~\cite{mcclean2018barren}. In other words, the variation in the cost function is concentrated in a shrinking region of parameter space, known as narrow gorge \cite{cerezo2021cost}. One notices the inverse relation between the spread of the Fourier coefficients and the size of the gorge. This is a manifestation of the uncertainty principle in Fourier analysis as first mentioned in Ref.~\cite{Enrico_2}. It stands to reason that the non-local connectivity of the ansatz due to the graph structure is sufficient to cause barren plateaus in an otherwise shallow and expressivity-restricted circuit. We have found this to hold true even for unweighted graphs, although random coefficients appear to amplify and accelerate the phenomenon.

We can see that values of TV decrease as we increase the number of qubits, while values of Fourier Density increase. This is a good example of why having several complementary metrics can be useful. 
In this particular case, barren plateaus seem to be characterized by very low Total Variation values and very large Fourier Density.

\subsection{Revisiting the initial problem}

At the end of this work, we wanted to apply the insights gained from the previous cases to the examples presented in the introduction. Upon inspecting the Hamiltonians, we can make some educated guesses about the cost function landscapes and their Fourier spectrum.
For both Hamiltonians, the resulting cost function has the period of $2 \pi$ in $f_{\gamma}$, which simplifies the comparison. 

\begin{figure}
    \centering
    \includegraphics[scale=0.24]{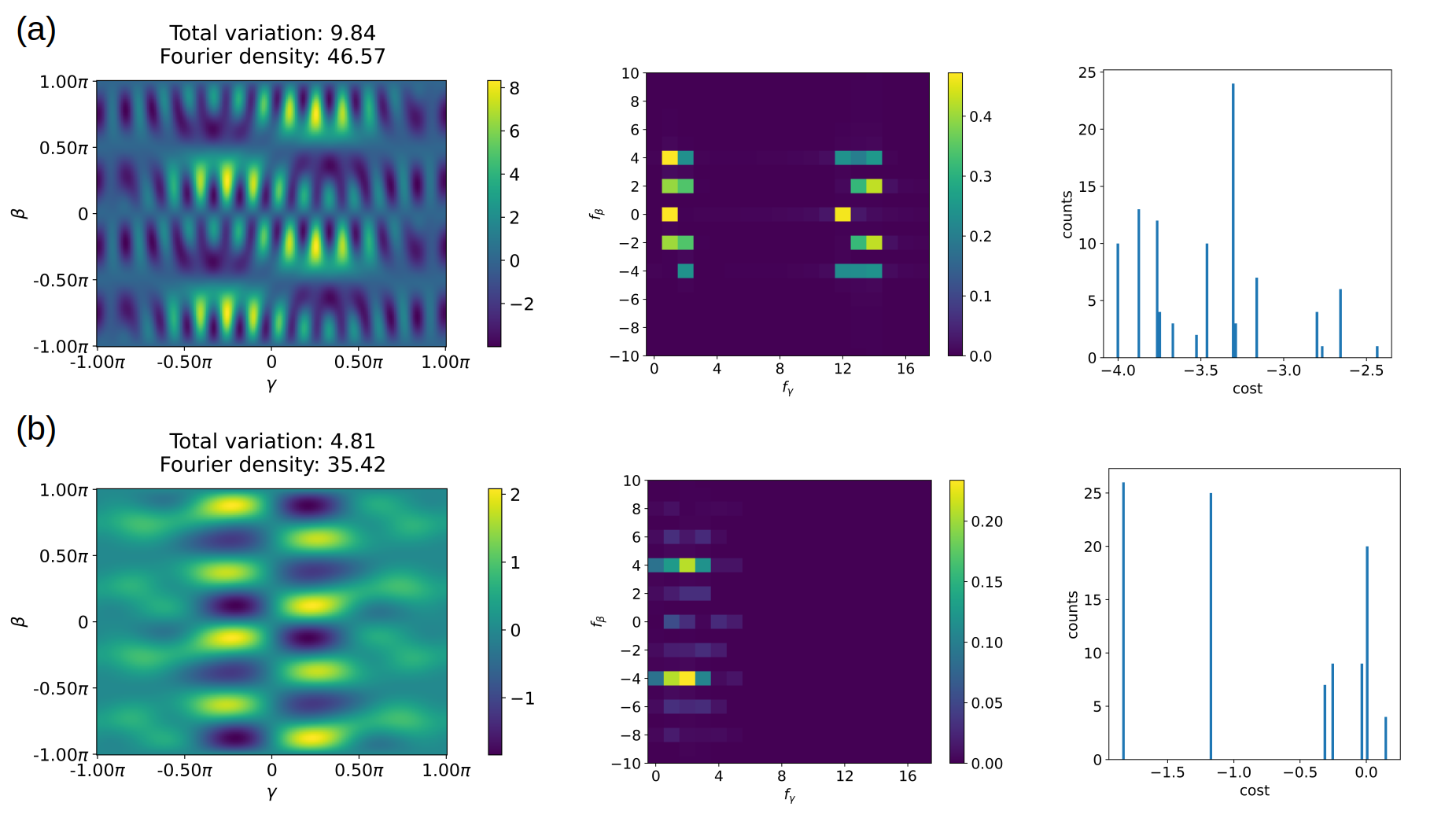}

\caption{Cost and Fourier landscapes for the Hamiltonians $H_1$ (a) and $H_2$ (b) as defined in Eqs. ~\ref{eq:H1} and ~\ref{eq:H2}, as well as histograms on final optimization outcomes. In accordance with the measured landscapes and recorded roughness metrics, not only is the optimizer able to find the optimal solution more than twice as often for $H_2$, but the density of sub-optimal local minima is significantly higher for $H_1$.}

\label{fig:initial_revisited}
\end{figure}

The first factor we analyze is whether high coefficients in the Hamiltonian lead to higher frequencies. $H_1$ has larger coefficients than $H_2$, and given that both have the same periodicity, we would expect to see points with larger frequencies $f_{\gamma}$ in the Fourier spectrum, which in turn cause fast oscillations in the $\gamma$ direction. This prediction is indeed correct, as we can see in Fig. \ref{fig:initial_revisited}.


The second factor is the order of the terms. Given that $H_2$ has 4-body terms, we expect to see $f_{\beta}$ up to $\pm 8$, while for $H_1$ they will be up to $\pm 4$. Consequently, cost function oscillations in the $\beta$ direction should be faster for $H_2$. However, while the maximal visible frequencies in Fig.~\ref{fig:initial_revisited} agree with the prediction, the dominant frequencies in terms of their amplitudes are on the two-body frequencies, i.e. $f_\beta = \pm 4$. This can be explained by looking at the relatively large coefficients of those terms in the Hamiltonian. It again appears that the magnitude of coefficients has a strong impact, because they not only directly control the $f_\gamma$ frequencies, but they indirectly affect which $f_\beta$ frequencies shape the cost landscape the most.

The effect of the structure is much more difficult to quantify in this case. $H_1$ consists of only 3 terms (all possible for two qubits), which makes it is hard to talk about a particular structure. On the other hand, when analyzing the structure of $H_2$ we can see that it has certain regularities. All terms have coefficients that are multiples of $\pm \frac{1}{8}$, all the 3-body terms are constructed in a similar way, and more. This may hint that certain frequencies will cancel, which would lead to a less rough landscape. However, this is speculative, as our work does not provide tools to assess any given Hamiltonian structure in-depth.


Sadly, we see that our work does not provide all the necessary tools to predict with certainty which of any two Hamiltonians produces a rougher landscape. It does, however, provide tools that allow us to quantify the roughness and to better understand which features of those Hamiltonians influence the shape of cost and Fourier landscapes. As we have shown in the previous section, our visualization techniques and roughness metrics are very helpful in understanding each particular contribution to the QAOA landscapes, but the sum of all components gives rise to a complexity that we do not know whether it can be predicted without a known reference landscape. 

\medskip

In the introduction, we suggest that the landscape produced by $H_1$ is more difficult to optimize compared to $H_2$, and that roughness is correlated with the performance of the optimizer. While quantifying this effect extensively is beyond the scope of this work, we test our hypothesis on this pair of Hamiltonians.  We performed 100 optimizations using the L-BFGS-B optimizer \cite{2020SciPy-NMeth,LBFGSB}. The initial parameters were drawn from a uniform distribution from [-$0.9 \pi$, $0.9 \pi$]. As demonstrated in Fig. ~\ref{fig:initial_revisited}, the optimization algorithm exhibits a higher likelihood of becoming trapped in local minima for $H_1$, which is in line with the higher yielded values for both roughness metrics. Additionally, we witness a higher density of low-lying local minima, which can be both a curse and a blessing. If the absolute global minimum needs to be found, these local minima present a big challenge because gradients will most likely point the optimizer towards one of them. On the flip side, if a \textit{good} solution suffices, which of the local minima one converges to may matter less than in a case similar to $H_2$, where any sub-optimal local minimum is significantly higher in energy.

\section{Conclusion}\label{Section:Conclusion}

This work was performed with the intention of understanding the relationship between a Hamiltonian and its QAOA cost landscape. 
Towards this goal, we applied visualization methods and utilized the Fourier transform of cost landscape scans. We also introduced roughness metrics, which quantify apparent changes in the landscapes and have been shown to correlate with optimization performance in classical applications~\cite{Wu2021UnderstandingLL, Aldeghi2022}. These tools enabled us to examine each component in a Hamiltonian of interest  and how it affects the cost function landscape.

Among the collection of metrics proposed by us, we conclude that Total Variation is the best option, as its outcome closely resembles a subjective interpretation of the cost landscape roughness. On the other hand, Fourier density exhibited hypersensitivity to some features of the cost function, such as the presence of fractional coefficients, but also to emergent properties in barren landscapes, to name a positive example. How correlated these metrics are with practical optimization performance remains unknown and is subject to future work. 

The visualization of the Fourier spectrum proved to be a useful tool in our study. It enabled a more thorough understanding of specific phenomena, such as the influence of high-order terms in the Hamiltonian, or the effect of their coefficients. It also simplified the process of comparing cost landscapes. We recommend it as a valuable tool to conduct analysis on cost function landscapes. This capability has since been incorporated into the \textit{orqviz} visualization package to make it available for use by other researchers in the field.

Finally, which properties of the Hamiltonian most affect the roughness? We have identified three properties of the Hamiltonian that influence the shape of the cost landscapes significantly. First are the relative magnitudes of the coefficients of the Hamiltonian terms. In general, larger relative magnitudes of the coefficients lead to rougher cost landscapes. 
Specifically, the presence of a single term with a substantially larger coefficient can significantly increase the roughness due to over-emphasizing a few frequencies in the landscape. 
The second influencing factor is the existence of higher-order terms in the Hamiltonian, i.e., operators which act on many qubits. 
Their presence increases the maximum present frequencies $f_\beta$ in the Fourier spectrum. Depending on the interplay with the coefficients, this can lead to a landscape with an increased number of saddle points and local minima.  
However, this increase in complexity can be counteracted by the third factor-- the structure of the Hamiltonian. 

Hamiltonians exhibiting certain structures seem to have lower roughness scores than those in which a structure has been disrupted by introducing or removing terms. Additionally, Hamiltonians with structures (which can potentially be traced back to the structure of a graph problem) appear robust to small perturbation. This phenomenon is clearly exemplified by the phenomenon of parameter concentration. This interestingly opens up the opportunity for studying entire families of practical QAOA Hamiltonians if they share a fundamental underlying structure. Given that the family of Hamiltonians studied in Sec.~\ref{sec:concentration_and_bps} are random 3-regular graphs, it is unclear how prevalent barren plateaus will be in more realistic cases of structured problems.



Factors that we have not found to play a significant role in increasing the roughness of the landscape are the number of qubits or the number of terms in the Hamiltonian. The latter is oftentimes linked to the former, but an increase in the number of qubits can, in fact, be shown to not only result in concentrating landscapes but also barren plateaus. They diminish the magnitude of smaller features and subsequently decrease the roughness of the cost function landscapes, irrespective of the precise number of qubits or individual terms.

It is important to note that there are limitations to the results presented in this study. 
First of all, we performed our study for QAOA with one layer. This limitation may affect the applicability of our findings to more practical scenarios. Visualizing and analyzing multidimensional cost functions presents a greater degree of difficulty. Fortunately, the \textit{orqviz} visualization library and the roughness metrics introduced in this work are designed for high-dimensional cost functions. 

Furthermore, we did not thoroughly study the relationship between roughness metrics and the performance of various optimizers, nor did we present a study on predicting the roughness of cost landscape for a given Hamiltonian. Finally, it appears valuable to perform a similar analysis on other variational algorithms, in which the Hamiltonian does not dictate the structure of the circuit. In such cases, the choice of the circuit ansatz plays a crucial role the shape of the optimization landscape. Notably, the approaches used in this work appear to be good candidates to study phenomena related to barren plateaus.

We believe that the development of tools to aid in ``opening up the black box'' that variational quantum algorithms currently are will facilitate the creation of more efficient algorithms in the future. There is significant value in the ability to examine a given problem from multiple perspectives, and the utilization of spectral analysis and metrics offers an alternate viewpoint to the most widely-used methods. This approach allowed for the identification of new insights, which to the best of our knowledge, have not yet been reported in the literature. 


\begin{acknowledgments}
The authors would like to thank Pierre-Luc Dallaire-Demers for suggesting to use Fourier transformation to analyze the landscapes of QAOA cost function to investigate the VQF Hamiltonians and to
Vladimir Vargas-Calderón for helpful comments on the manuscript.
The authors would also like to acknowledge the \mbox{ORQUESTRA\textsuperscript{\textregistered}}\ platform by Zapata Computing Inc. that was used for collecting the data presented in this work.

\end{acknowledgments}

\bibliographystyle{unsrt}
\bibliography{bibliography}

\appendix


\section{Visualization methods}\label{sec:viz-methods}

For visualization purposes, we used the \textit{orqviz} package~\cite{orqviz}, which is a Python library for visualizing the cost function landscapes of parametrized algorithms. While \textit{orqviz} offers more sophisticated visualization techniques, we utilize it only to easily perform 2D scans of the QAOA cost landscapes, which can then be Fourier-transformed into their corresponding Fourier landscape. The latter is a functionality which we added to \textit{orqviz} during this project. 

\subsection{Cost Landscape}

To visualize the cost landscape, we perform 2D grid scans of the cost function with a $201 \times 201$ resolution, centered around the parameter values $(\beta, \gamma) = (0, 0)$ (except for Fig. \ref{fig:single_big}, where we used $401 \times 401$. The basis vectors for the 2D grid can in principle be chosen to be any two (preferably orthogonal) parameter vectors. However, since we restrict ourselves to $p=1$ QAOA with two free parameters, the vectors are chosen in the  $\gamma$ and $\beta$ directions. We decided to use $\pi$ as the extent for both axes for most of the figures presented here, which simplifies the analysis and comparison.

\subsection{Fourier Landscape}
\label{sec:vis_fourier}
To visualize the Fourier space, we perform a discrete fast Fourier transform (FFT) on a 2D scan of the cost function landscape. To illustrate how to interpret the output of the Fourier transform, we will use the following simple function: $\mathcal{C}(\beta, \gamma)=\cos(x\gamma)\cos(y\beta)$. When we apply the Fourier transform using eq. \ref{eq:main-ft}, it produces a point with coordinates $\omega=(f_\gamma, f_\beta) = (x, y)$ (as well as points at the corresponding negative frequencies). Therefore the frequencies associated with $\gamma$ and $\beta$ are being equal to $f_{\gamma} = x$ and $f_{\beta} = y$ accordingly (see Sec. \ref{sec:fourier}).

In our experiments, we plot only the absolute magnitude of the discrete Fourier transform's complex output.
The magnitude of a coefficient determines the amplitude of its corresponding frequency component and its influence on the landscape.
On the other hand, the phase dictates the location of crests and troughs of frequencies relative to one another. 
For instance, $\sin(x)$ and $\cos(x)$, which are just shifted versions of each other, both have coefficients of magnitude $1/2$ but with different phases.
The analysis of various plots has confirmed that the phases do not provide any significant insight, so we decided not to include this in the data presented here. 

As described in Sec. \ref{subsec_results:toy}, we decided to plot only half of the Fourier spectrum, by cutting out the two quadrants corresponding to $f_\gamma < 0$. To summarise, the reason is that half of the spectrum is redundant due to the cost function being real, which imposes a twofold symmetry in the coefficients $c_{(f_\beta, f_\gamma)} = c^*_{(-f_\beta, -f\gamma)}$.
Finally, to ensure that plots are easy to compare, we remove the constant offset from the visualization, setting $c_{(0, 0)} = 0$.

\subsubsection{Choosing period and maximum frequency}

\begin{figure}
    \begin{center}
        \includegraphics[width=1\textwidth]{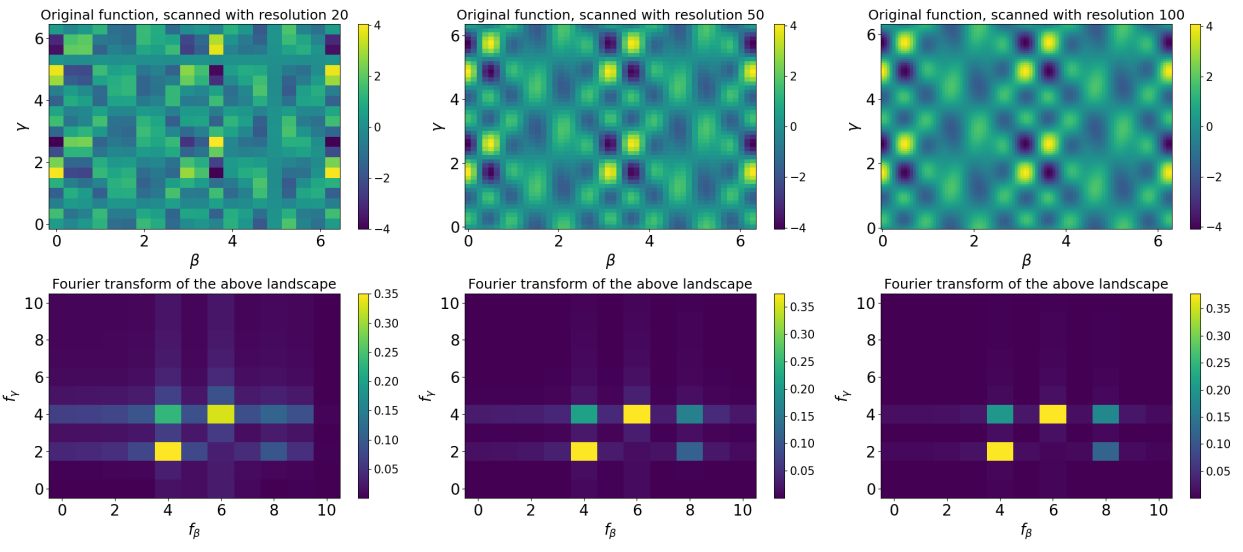}
        \caption{Impact of cost landscape scan resolution (top row) and the respective Fourier spectrum (bottom row).
        The scan resolutions are $20 \times 20$ (a), $50 \times 50$ (b) and $100 \times 100$ (c).
        Increasing the resolution of the cost landscape scans results in a reduction of artifacts in the Fourier spectrum, revealing the true frequencies.}
        \label{fig:resolution_artifact}
    \end{center}
\end{figure}

Before producing a Fourier plot, one must undertake some necessary considerations. Firstly, it is important to ensure that the 2D scan grid resolution is high enough when the intention is to perform a FT of the landscape. Insufficient scan resolution causes spurious frequencies to appear in the Fourier spectrum, a phenomenon known as ``aliasing''.
This behaviour is very noticeable in lower resolution scans (linear resolution $<100$) and will additionally inflate the value of the Fourier density roughness metric. The nonzero artifacts tends to extend outwards from true frequencies like streaks as shown in Fig.~\ref{fig:resolution_artifact}. 
Secondly, one must make sure that at least one period of the function is scanned, otherwise, the key assumption of periodicity does not hold, and taking a discrete Fourier transform does not return a faithful representation of the spectrum.

Both the minimum resolution and the extent of the scans should depend directly on the Hamiltonian that generates the rotations, i.e., the cost Hamiltonian. Since the frequencies in the cost function are the pairwise differences of the Hamiltonian's eigenvalues, the minimum recommended scan resolution to resolve landscape features is given by the highest frequency, i.e., the largest difference of eigenvalues.
Vice versa, the period is obtained from the greatest common denominator (GCD) of the frequencies. Writing this as $GCD(\{\omega\})$, the period is $\frac{2\pi}{GCD(\{\omega\})}$. This is because the period is $2\pi$ times the smallest number that multiplies all frequencies to integers, which is $\frac{1}{GCD(\{\omega\})}$. Since the frequencies are the differences of eigenvalues, the GCD of the eigenvalues will be a divisor of the GCD of the frequencies.
For the case of diagonal Hamiltonians such as the QAOA problem Hamiltonian, the eigenvalues are themselves sums and differences of the Hamiltonian coefficients. Take an Hamiltonian defined as
\begin{equation}
    H = \sum_{k=1}^m c_k H_k
\end{equation}
where the $H_i$ are diagonal Pauli strings (Pauli strings composed of $I$ and $Z$) with eigenvalues $\pm 1$. Then one can see that the eigenvalues of $H$ obey $\lambda_i(H) \in \{\pm c_1 \pm c_2 \cdots \pm c_m\}$, meaning that the frequency spectrum when this Hamiltonian is used as generator will be
\begin{equation}
    \label{eq:max_spect}
    \Omega_\gamma = \{\lambda_i(H) - \lambda_j(H)\}_{i,j} \subset \{\pm \{2c_1, 0\} \pm \{2c_2, 0\} \dots \pm \{2c_m, 0\}\} .
\end{equation}

\FloatBarrier
Therefore for QAOA, a suitable period in the $\gamma$ direction can be found from the GCD of the coefficients of $H$, without the need to know the eigenvalues. Since the $c_k$'s always appear in the spectrum in multiples of 2, the period is at most
$\frac{\pi}{GCD(\{c_k\})}$.
Regarding the maximum theoretical frequency, which informs the sampling resolution, we can use Eq.~\ref{eq:max_spect} to infer that this is
\begin{equation}
    \max|f_\gamma| \le 2\sum_k |c_k|.
\end{equation}

Conversely, for the $\beta$ direction the generator is always $\sum_{i=1}^n X_i$, which gives a theoretical frequency spectrum
\begin{equation}
    \Omega_\beta \subset \{0, \pm 2, \pm 4, \cdots \pm 2n\}.
\end{equation}
The period is therefore simply $\pi$. The maximum frequency in theory is $2n$, however for $p=1$ QAOA one finds (by backpropagating the measurement Hamiltonian in the Heisenberg picture) that in practice
\begin{equation}
    \max|f_\beta| \le 2w,
\end{equation}
where $w$ is the maximum weight of the Pauli strings composing the problem operator.

\section{Theoretical frequencies of selected Hamiltonians}\label{secA1}

In this appendix, we will show how we arrived at the theoretical frequencies for Hamiltonians of form $Z_0 + Z_1 + cZ_0Z_1$ presented in Table~\ref{table:toy-problem-frequencies}. As explained in the text, for brevity we write the absolute values of the frequencies, as for each positive frequency the corresponding negative frequencies are also present. We will start with equations for linear terms ($Z_0$, $Z_1$), derived from Eq. (\ref{eq:toy-problem}).

\begin{equation} \label{eq:linear-terms}
\begin{aligned}
\langle C_0 \rangle = \langle C_1 \rangle &= \sin(2\beta)\sin(2\gamma)\cos(2c\gamma) \\
& = \frac{1}{2}\sin(2\beta)[\sin((2c+2)\gamma)+\sin((2c-2)\gamma)]
\end{aligned}
\end{equation}

In Eq.~\eqref{eq:linear-terms}, the single $\beta$-containing factor $\sin(2\beta)$ means that $f_{\beta}=2$. To find $f_{\gamma}$, we can use the identity $\sin(x)\cos(y) = \frac{1}{2} [\sin(x + y) + \sin(x - y)]$. This tells us that $\sin(2\gamma)\cos(2c\gamma)$ produces both $f_{\gamma} = 2c + 2$ and $f_{\gamma} = 2c - 2$. Combined, the (absolute value) frequencies produced by linear terms are $(2c\pm2, 2)$.

Exception to this rule: When $c=1$, $\langle C_{0} \rangle=\frac{1}{2}\sin(2\beta)[\sin(4\gamma)+\sin(0\cdot\gamma)]=\frac{1}{2}\sin(2\beta)\sin(4\gamma)$ as given by Eq.~\eqref{eq:linear-terms}. Contrary to the previous prediction, in this case the only frequency left is (4, 2). The component (0,2), predicted to appear in the term $\sin(2\beta)\sin(0\gamma)$, does not show up because $\sin(0\cdot\gamma)=0$ and therefore the entire term is equal to 0.

\begin{align}
\langle C_{01} \rangle  &= c \sin(4\beta) \sin(2c\gamma) \cos(2\gamma) - \frac{c}{2} \sin(2\beta)^2 [\cos(4\gamma) - 1] \nonumber\\
& = \frac{c\sin(4\beta)}{2}[\sin((2c+2)\gamma)+\sin((2c-2)\gamma)] - \frac{c}{2}  [-\frac{\cos(4\beta)}{2}+\frac{1}{2}] [\cos(4\gamma) - 1] \nonumber\\
& = \frac{c}{2}[\sin(4\beta)\sin((2c+2)\gamma)+\sin(4\beta)\sin((2c-2)\gamma)] \; \nonumber\\
& \hspace{0.5cm} + \frac{c}{4} [\cos(4\beta)\cos(4\gamma)+\cos(4\gamma)-\cos(4\beta)+1]\,. \label{eq:quadratic-terms}
\end{align}

In Eq.~\eqref{eq:quadratic-terms}, each term represents one distinct frequency component. In the respective order of terms, the frequencies $\omega=(f_{\gamma}, f_{\beta})$ contributing to $\langle C_{01} \rangle$ are $(2c+2, 4)$, $(2c-2, 4)$, (4, 4), (4, 0), (0, 4), and (0, 0). Note that terms missing a $\cos(\gamma)$ or $\cos(\beta)$ component can be rewritten to include terms with a frequency of $0$, for instance $\cos(4\gamma)=\cos(0\beta)\cos(4\gamma)$ and $1=\cos(0\beta)\cos(0\gamma)$. The trigonometric identities $\sin^2(x)=\frac{1-\cos(2x)}{2}$ and $\sin(x)\cos(y)=\frac{1}{2}[\sin(x+y)+\sin(x-y)]$ were used in the simplification of the above equations.

\section{Metrics}\label{secA2}

We experimented with several different roughness metrics before deciding on the two metrics we ultimately used for our research (which are explained in Section \ref{sec:roughness-metrics}). In this appendix, we discuss these other metrics and explore their strengths and weaknesses.

\subsection{Total variation family}

We first attempted calculating TV using an algorithm defined in Section 3 of Ref.~\cite{Wu2021UnderstandingLL}, which we explained in Section~\ref{sec:background_total_variance}). In that work, TV is scaled by a factor of $\frac{1}{b-a}$ such that its value is invariant to scaling of the function's domain. In the case of a periodic function, if one chooses the sampling domain such that it covers exactly one period of the cost function, this normalization can be omitted. To estimate the TV, we sample 200 random directions, with 200 samples taken along each of these directions, which usually gives an approximate std$/$mean ratio $\leq10\%$ between runs. We found this ratio a good compromise between the computation time and required accuracy.

In Ref.~\cite{Wu2021UnderstandingLL}, the final roughness index is defined as:

\begin{equation}\label{eq:stdev-over-mean}
I := \frac{\sigma}{\mu}
\end{equation}

where $\mu$ and $\sigma$ are the mean and standard deviation of normalized TV in different directions. This roughness index measures how much TV \emph{varies in different directions} while seeming unaffected by the raw value of TV. Experimentally, we found that that landscapes with higher $\mu$ also tend to have higher values of $\frac{\sigma}{\mu}$. Since calculating $\mu$ is simpler and easier to interpret than $\frac{\sigma}{\mu}$, we decided to use only it. However, this choice might depend on the type of the landscapes analyzed, since for some landscapes studied in Ref.~\cite{Wu2021UnderstandingLL}, while the value of $\mu$ was constant, the value of $\sigma$ was changing significantly, so it might be valuable to monitor both values in future experiments.

In addition, we experimented with another version of estimating TV by numerically approximating the $m$-dimensional integral of the cost function (where $m$ is the number of parameters) through full discrete integration. Especially on low dimensions, this provides similar results to using our randomized estimation of $\mu$ and does not require setting metaparameters such as the number of random directions. However, the sampling requirement scales exponentially with the number of dimensions, while estimating $\mu$ doesn't require more computation time as the number of dimensions increases, as long as the number of sampled directions is fixed. This did not matter in the context of this work, but we still decided to use $\mu$, as a potentially more future-proof metric.

\subsection{Fourier metric family}

As discussed in Ref.~\ref{sec:background_fourier_density}, large Fourier coefficients on higher frequencies mean that a function is more oscillatory and hence should have a higher roughness index. As observed in Ref.~\cite{siddiqi1972order}, the amplitudes that large frequencies can have are bounded by total variation, which provides further evidence that roughness metrics in the Fourier space can directly quantify roughness of the cost landscape, but may contain complementary information. It holds that


\begin{equation}\label{eq:fourier-bounds}
	\lvert \omega \rvert c_{\omega} \leq \frac{2}{\pi} TV(\mathcal{C}),
\end{equation}

where $\omega$ represents the frequency, $c_{\omega}$ is the coefficient at that frequency, and $\mathcal{C}$ is the cost function.
Inspired by the relation in Eq.~\eqref{eq:fourier-bounds}, we explore the following quantity as a potential metric
for the roughness of a multidimensional function:
\begin{equation}\label{eq:fourier-max-equation}
R_f^{\text{max}} := \max_{\omega}[\lvert c_{\omega} \rvert \cdot\|\omega\|]\,,
\end{equation}
where we are now considering a $n$-dimensional Fourier transform and thus the coefficients are labeled by a frequency vector $\omega \in \mathbb{Z}^N$. Up to constant factors, this can be seen to upper bound the TV.
We refer to this quantity as ``Fourier maximum''.

When the metric is defined like this, however, it is sensitive to small changes in the value of the high-frequency coefficients and is blind to the low-frequency modes.
We, therefore, experimented with the index obtained by taking the mean over the Fourier coefficients instead of the maximum,

\begin{equation}\label{eq:fourier-sum-equation}
	R_f^{\text{mean}} := \frac{1}{\lvert \hat{\omega} \rvert}\sum_{\omega}\lvert c_{\omega} \rvert \cdot\|\omega\|\,,
\end{equation}
where $\hat{\omega}$ is the maximal present frequency.
We refer to this metric as ``Fourier mean''.
We observe that Fourier density (described in Sec. \ref{sec:background_fourier_density}) and Fourier mean correlate very well with each other. However, we decided to use Fourier Density due to its numerical stability and ease of computation (see \ref{sec:background_fourier_density}).  It also does not come with the issues that Fourier maximum has and therefore we decided to use it as the most reliable metric from this family.



\end{document}